\documentclass[reprint,superscriptaddress,aps,prb,twocolumn,floatfix]{revtex4-2}
\usepackage{setspace}
\usepackage{amsmath}
\usepackage{bm}
\usepackage{graphicx}
\usepackage[nearskip,margin = 0pt]{subfig}
\usepackage{verbatim}
\usepackage{amsfonts}
\usepackage{amssymb}
\usepackage{textcomp}
\usepackage{mathrsfs}
\usepackage{mathtools}
\usepackage{url}
\usepackage{caption}

\usepackage{xcolor}
\usepackage[colorlinks,linkcolor=blue,anchorcolor=blue,citecolor=blue,urlcolor=black]{hyperref}
\usepackage{ragged2e}
\usepackage{float}
\usepackage{lineno}
\DeclareGraphicsExtensions{.pdf,.eps,.png,.jpg,.mps}
\begin{document}
%\pagewiselinenumbers

\title{Bridging microcombs and silicon photonic engines for optoelectronics systems}
\author{ Haowen Shu$^{1,\dagger}$, Lin Chang$^{2,\dagger}$, Yuansheng Tao$^{1,\dagger}$, Bitao Shen$^{1,\dagger}$, Weiqiang Xie$^{2}$, Ming Jin$^{1}$, Andrew Netherton$^{2}$, Zihan Tao$^{1}$, Xuguang Zhang$^{1}$, Ruixuan Chen$^{1}$, Bowen Bai$^{1}$, Jun Qin$^{1}$, Shaohua Yu$^{1,3}$, Xingjun Wang$^{1,3,4,*}$ and John E. Bowers$^{2,*}$\\
\vspace{3pt}
$^1$State Key Laboratory of Advanced Optical Communications System and Networks, Department of Electronics, School of Electronics Engineering and Computer Science, Peking University, Beijing 100871, China.\\
$^2$Department of Electrical and Computer Engineering, University of California, Santa Barbara, CA 93106, USA.\\
$^3$Peng Cheng Laboratory, Shenzhen 518055, China.\\
$^4$Frontiers Science Center for Nano-optoelectronics, Peking University, Beijing 100871, China.\\
$^\dagger$These authors contributed equally to this work\\
\vspace{3pt}
Corresponding authors: $^*$xjwang@pku.edu.cn, $^*$bowers@ece.ucsb.edu.}

%\begin{abstract}

%\end{abstract}

\date{\today}

\maketitle
\noindent
\large\textbf{Abstract} \\
\normalsize\textbf{Microcombs have sparked a surge of applications over the last decade, ranging from optical communications to metrology. Despite their diverse deployment, most microcomb-based systems rely on a tremendous amount of bulk equipment to fulfill their desired functions, which is rather complicated, expensive and power-consuming. On the other hand, foundry-based silicon photonics (SiPh) has had remarkable success in providing versatile functionality in a scalable and low-cost manner, but its available chip-based light sources lack the capacity for parallelization, which limits the scope of SiPh applications. Here, we bridge these two technologies by using a power-efficient and operationally-simple AlGaAs on insulator microcomb source to drive CMOS SiPh engines. We present two important chip-scale photonic systems for optical data transmissions and microwave photonics respectively: The first microcomb-based integrated photonic data link is demonstrated, based on a pulse-amplitude 4-level modulation scheme with 2 Tbps aggregate rate, and a highly reconfigurable microwave photonic filter with unprecedented integration level is constructed, using a time stretch scheme. Such synergy of microcomb and SiPh integrated components is an essential step towards the next generation of fully integrated photonic systems.
}

\vspace{3pt}
\maketitle
\noindent
\large\textbf{Introduction} \\
\normalsize
\noindent Integrated photonics is profoundly impacting data communication and signal processing \cite{PIC1,PIC2,PIC3,PIC4,PIC5}. A crucial development in the last decade is the demonstration of Kerr microcombs, which provide mutually coherent and equidistant optical frequency lines generated by microresonators \cite{DKS_RV1,DKS_RV2,DKS_RV3}. This coherent source is a valuable tool in applications such as microwave photonics \cite{MP1,MP2,MP3}, precise optical frequency synthesizers \cite{OFS1} and optical atomic clocks \cite{OAC1,OAC2}. Meanwhile, their multi-wavelength output enables high data-processing capability in parallelized optical communication \cite{OC1,OC2,OC3}, computation \cite{Cpt1,Cpt2} and sensing \cite{Ss1,Ss2,Ss3}. Therefore, microcombs hold the promise to extend the application space of integrated photonics to a much broader scope. So far, however, in almost all system-level-demonstrations leveraging microcomb technologies, the passive comb generator is the only component that is integrated on-chip. The rest of the system, including the comb pumping lasers, passive and active optical components and the supporting electronics, usually rely on bulky, expensive and power-hungry equipment, thereby undermining the promised benefits of integrated photonics.

 \begin{figure*}[ht]
\centering
\captionsetup{singlelinecheck=no, justification = RaggedRight}
\includegraphics[width=18cm]{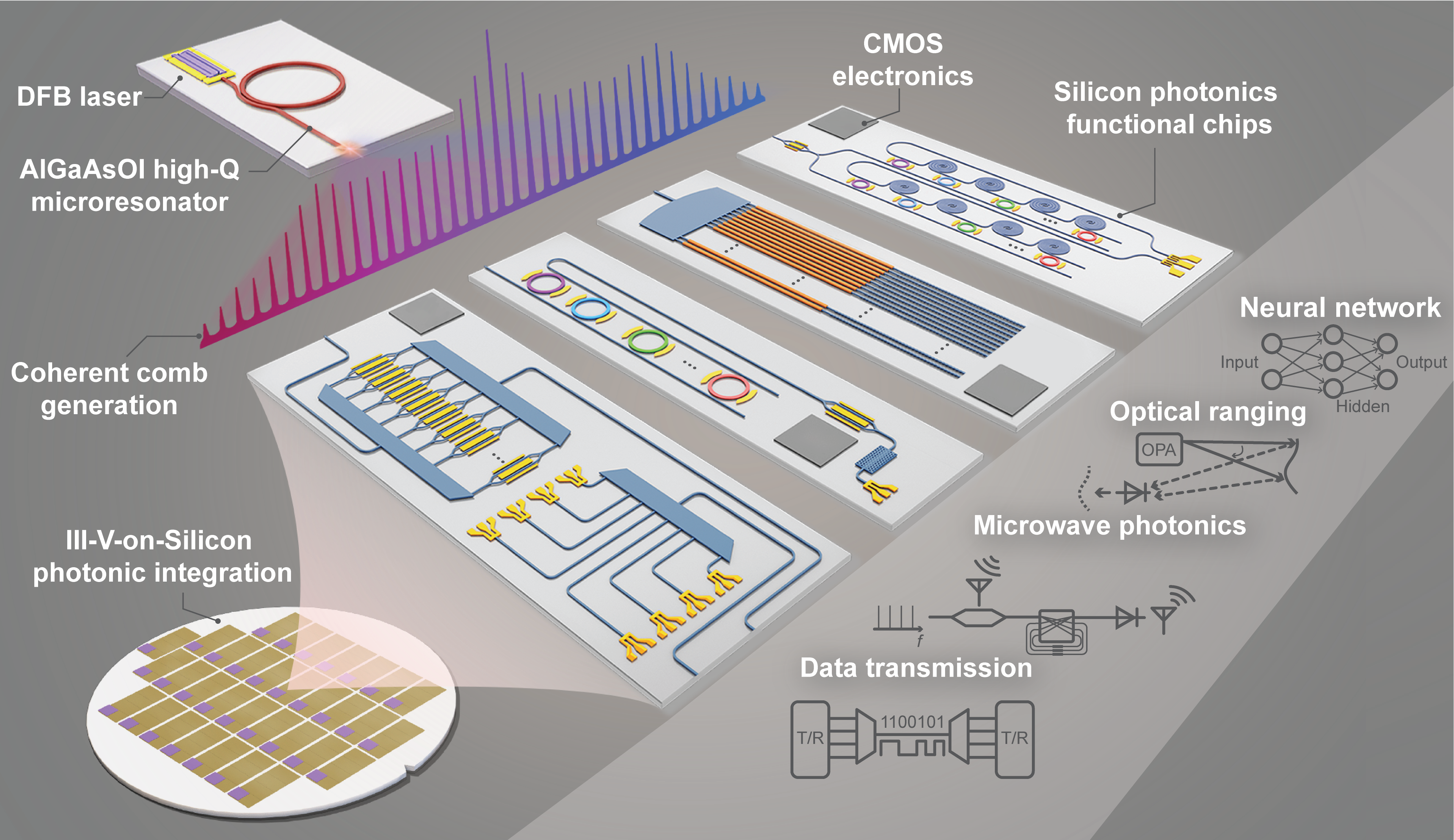}
\caption{\textbf{Microcomb based silicon photonics optoelectronics systems.} Conceptual drawings for several integrated optoelectronics systems (\textbf{i}, data transmission, \textbf{ii}, microwave photonic signal processing, \textbf{iii}, optical beam steering and \textbf{iv}, photonic computing) realized by combining a microcomb source with silicon photonic chips.
}
\label{fig1}
\end{figure*}
On the other hand, the advances in silicon photonics (SiPh) technology have provided an excellent solution in miniaturization of optical systems \cite{SiPh1,SiPh2,SiPh3,SiPh4,SiPh5,SiPh6,SiPh7,SiPh8,SiPh9}. Benefiting from CMOS compatible manufacturing, SiPh has enabled scalable and low-cost integration of diverse functions, including lasers, modulators, photodetectors and passive interferometers as well as control electronics on one or multiple SOI chips \cite{SiPh4,SiPh5,SiPh6,SiPh7,SiPh8,SiPh9}. These ‘photonic engines’, have been commercialized in data interconnects \cite{SiPh_DT1,SiPh_DT2,SiPh_DT3}, and widely applied in photonic neural networks \cite{SiPh_Cpt1,SiPh_Cpt2}, microwave photonics \cite{SiPh_MP1,SiPh_MP2,SiPh_MP3}, and optical ranging \cite{SiPh_Rg1,SiPh_Rg2,SiPh_Rg3}. Yet, a key ingredient missing from foundry-based SOI PICs is the multiple wavelength source. For example, the current state-of-the-art photonic transceiver module contains an 8 channel Distributed Feedback Laser (DFB) array for wavelength division multiplexing (WDM) \cite{8DFB}. Increasing the channel count in such a system requires considerable design effort such as line-to-line spacing stabilization and increased assembly workload. Moreover, the lack of mutual coherence among channel lines restricts some applications, such as precise time-frequency metrology. 

Although uniting these two technologies seems an obvious way to address the aforementioned problems on both sides, until now, such a combination has remained elusive. Previously, while the combinations of a microcomb and other photonic components have shown potential in optical computation \cite{Cpt1}, atomic clocks \cite{OAC1} and synthesizer systems \cite{OFS1}, these integrated demonstrations have been limited to only one or a few separate devices relying on specialized fabrication processes which are not suitable for high-volume production. Moreover, comb startup \cite{Trig1,Trig2,Trig3} and stabilization techniques \cite{Stb1,Stb2,Stb3}, which requires high performance discrete optics and electronic components, dramatically increase the operation complexity and system size. Recent progress in hybrid or heterogeneous laser-microcomb integration enables on-chip comb generation in a simplified manner \cite{SIL1,SIL2,SIL3}, but these integration schemes add complexity in processing. 
These difficulties, along with the extra expenditures on multi-channel match-up and other pretreatments in system operations, have so far obstructed the implementation of a functional laser-microcomb system.

Here, we make a key step in bridging these two essential technologies. Using an AlGaAs-on-insulator (AlGaAsOI) microresonator that can be directly pumped by a DFB on-chip laser, a dark-pulse microcomb is generated, which exhibits state-of-the-art efficiency, simple operation and great longtime stability. Such a coherent comb is used to drive CMOS-foundry-based SiPh engines containing versatile functionality which can be used for a wide range of applications (see Fig. \ref{fig1}). Based on this approach, system-level demonstrations are presented for two major integrated photonics fields: (1) As a communications demonstration, we present 
the first microcomb-SiPh transceiver-based data link with 100 Gbps PAM4 transmission and 2 Tbps aggregate rate for datacenters, and (2) for microwave photonics, a compact microwave filter is demonstrated with tens-of-\textmu s-level speed by an on-chip multi-tap delay line processing scheme, whose tunable bandwidth and flexible center frequency are capable of supporting 5G, radar and on-chip signal processing. This work paves the way towards the full integration of a wide range of optical systems, and it will significantly accelerate the proliferation of microcomb and SiPh technologies for the next generation of integrated photonics.

%Fig2
\begin{figure*}[ht]
\centering
\captionsetup{singlelinecheck=no, justification = RaggedRight}
\includegraphics[width=18cm]{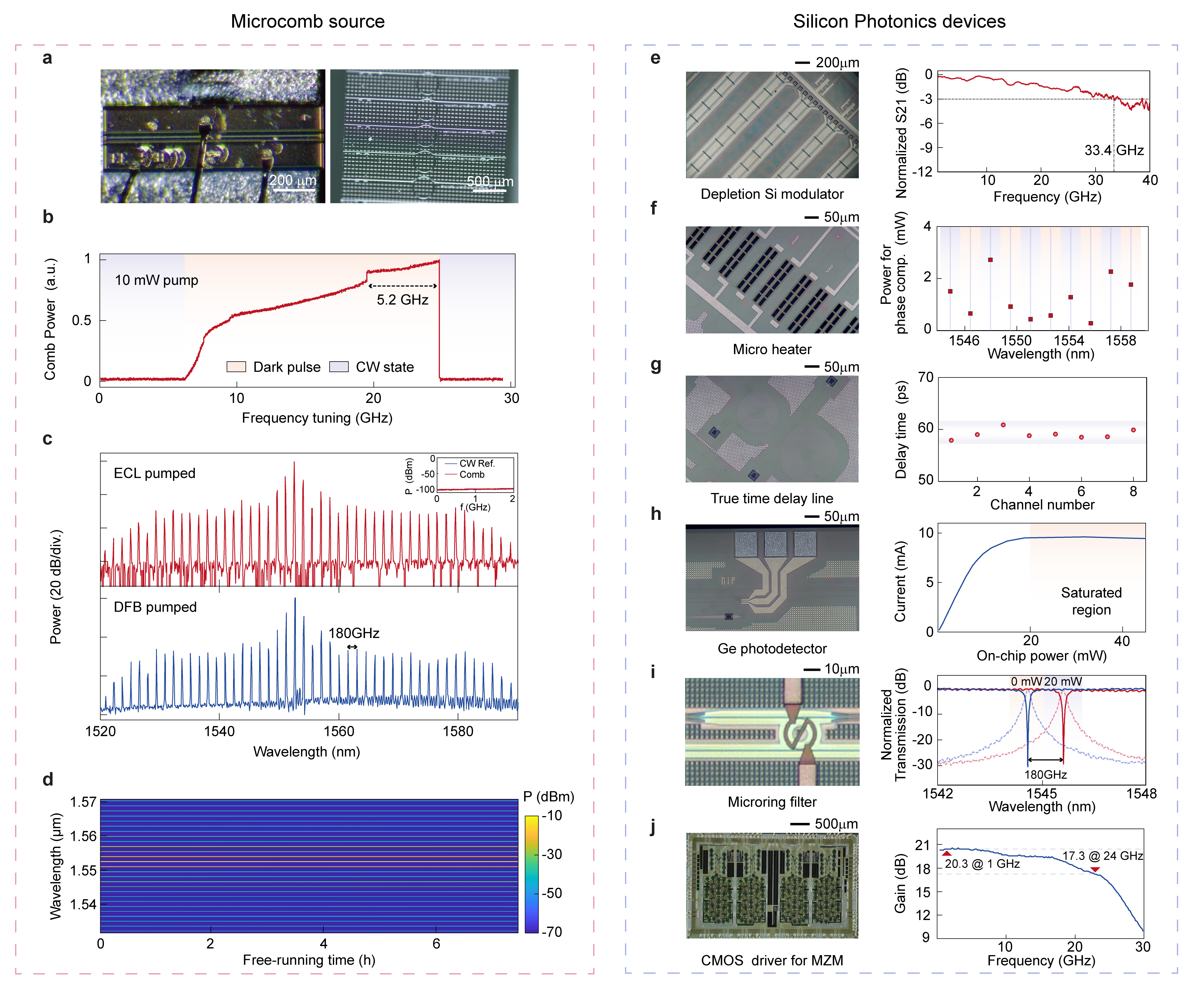}
\caption{\textbf{Comb generation and fundamental characteristics. a}, Optical image of the InP DFB laser chip and the AlGaAs-on-insulator microresonators for dark-pulse generation. The free spectral range (FSR) of the 144 \textmu m radius rings are about 90 GHz. 
\textbf{b}, Normalized comb power when tuning the pump frequency across the resonance at around 1552 nm. With 10 mW pump power, dark-pulse Kerr comb could be accessed in a large frequency window (tens of gigahertz). No power saltation occurs during the transition from continuous wave states to dark-pulse states, indicating the elimination of the well-known triggering problem in bright soliton generation.
\textbf{c}, 2-FSR dark-pulse spectra pumped by a commercial external laser or a DFB laser chip. Inset: comb intensity noise (resolution bandwidth 100 kHz). The intensity noise of the dark-pulse Kerr comb is at the same power level as the Electrical Spectrum Analyzer (ESA) background.
\textbf{d}, Long-term stability of a free running comb. The spectra are recorded by a high-resolution optical spectrum analyzer (OSA) every 5 minutes.
\textbf{e}, Depletion mode Si Mach-Zehnder modulators. The typical 3dB operation bandwidth is $>$ 30 GHz.  
\textbf{f}, MZ inteferometer based TiN micro heaters used for phase compensation. The resistance is approximately 200 ohm, with a phase tuning efficiency of $\sim$ 20 mW/$\pi$ . 
\textbf{g}, Silicon spiral waveguide for on-chip true time delay. Euler curves are used in the spiral waveguide for adiabatic bending. For a 60 ps delay line, the total loss is $<$ 0.5 dB, with delay time variation of $<$3\%. 
\textbf{h}, Vertical epitaxial Ge photodetectors. The responsivity declines with the increasing on-chip power. A saturated point could be reached when the power is further increased, causing a saturation power of $\sim$20 mW.
\textbf{i}, Microring filter as used in the on-chip wavelength selective switch. Frequency shift of 180 GHz can be obtained with 20 mW power dissipation. 
\textbf{j}, CMOS driver for signal amplification before injection into MZ modulator (not used in high bit rate ($>$50 Gbps) signal transmission experiment). The 3-dB gain bandwidth is $\sim$24 GHz.
}
\label{fig2}
\end{figure*}
%Numerous integrated nonlinear photonics platforms, such as SiN \cite{31}, AlN \cite{32}, Hydex \cite{33}, LiNbO$_3$ \cite{34}, AlGaAs \cite{35}, GaP \cite{36}, GaN \cite{37} and SiC \cite{38}, have been developed recently to implement a chip-scale Kerr resonator.

\vspace{6pt}
\noindent
\large\textbf{Building blocks}\\
\noindent
\normalsize
\textbf{AlGaAsOI microcombs}\\
\noindent
The integrated comb source employed for the chip-scale optical systems in this work is based on an AlGaAsOI platform using heterogeneous integration, as shown in Fig. \ref{fig2}a. This process is currently realized at the 100 mm wafer-scale without any strict fabrication processes such as CMP or high temperature annealing. It can therefore be directly adopted by current III-V/Si photonic foundries \cite{liang2021recent}. With the advanced fabrication process discussed in ref. \cite{AlGaAs1,AlGaAs2}, quality ($Q$) factor $>$ 2 million can be obtained in the AlGaAsOI resonator, corresponding to a waveguide loss $<$ 0.3 dB/cm. Combined with the extremely high third order nonlinear coefficient of AlGaAs (n2 = 2.6 $\times$ 10$^{-17}$m$^2$W$^{-1}$), Kerr comb generation from the AlGaAsOI microresonators (right panel of Fig. \ref{fig2}a) exhibits a record-low parametric oscillation threshold down to tens of \textmu W and coherent comb state generation under pump power at a few mW level, which can be satisfied by a commercial InP DFB laser chip (left panel of Fig. \ref{fig2}a).

A dark-pulse state \cite{DP1,DP2} is used in this work to achieve robust, high efficiency coherent microcombs\cite{algaasdark}. This state works in the normal dispersion regime with the assistance of the avoided mode crossing (AMX) effect (see Supplementary Note I). Compared with bright solitons working in the anomalous dispersion regime, the dark-pulse is inherently tolerant to thermal effects that usually make bright soliton states difficult to access \cite{Trig1}. Previously, thermal instability imposed a strict requirement on tuning and feedback control for bright soliton generations. Particularly for AlGaAsOI, whose thermo-optical effect (2.3 $\times$ 10$^{-4}$K$^{-1}$) is one order of magnitude higher than that of Si$_3$N$_4$ 
or Silica, cryogenic setups were needed to access soliton states \cite{Trig2}. In contrast, dark-pulse operation experiences a much smaller power step during the transition from the continuous wave state to the coherent comb state (see Supplementary Note II). More importantly, due to the thermo-induced self-stable equilibrium mechanism of microcavities, strong thermo-optic effects of AlGaAsOI here can be leveraged to significantly extend the accessibility window of the coherent comb state \cite{Tml1}.  Such behavior is experimentally characterized in Fig. \ref{fig2}b, where comb power versus pump frequency detuning is recorded, showing the accessible frequency range of the dark-pulse to tens of GHz-level, about 10 times wider than that in bright soliton \cite{Trig1}.

Together, the traits above make coherent comb generation efficient and robust in AlGaAsOI microresonators, with greatly simplified operation. Fig. \ref{fig2}c shows the dark-pulse spectra pumped by an external cavity laser and a DFB laser chip, respectively, with the same on-chip power of 10 mW. Low noise combs are generated by simply tuning the laser to a resonant frequency, with no extra active capture or dedicated tuning techniques (Fig. \ref{fig2}c). Moreover, benefiting from the self-stabilization enabled by the strong thermo-effect, the comb is able to maintain stable operation with no feedback loop. Fig. \ref{fig2}d shows the spectral power versus time in a free running AlGaAs dark-pulse, with little power fluctuations in $>$ 7 hours. The simplicity of both generation and stabilization facilitate seamless implementation of AlGaAsOI microcombs in current optoelectronic systems, making the system well-suited for practical applications.
 
%Fig3
\begin{figure*}[ht]
\centering
\captionsetup{singlelinecheck=no, justification = RaggedRight}
\includegraphics[width=18cm]{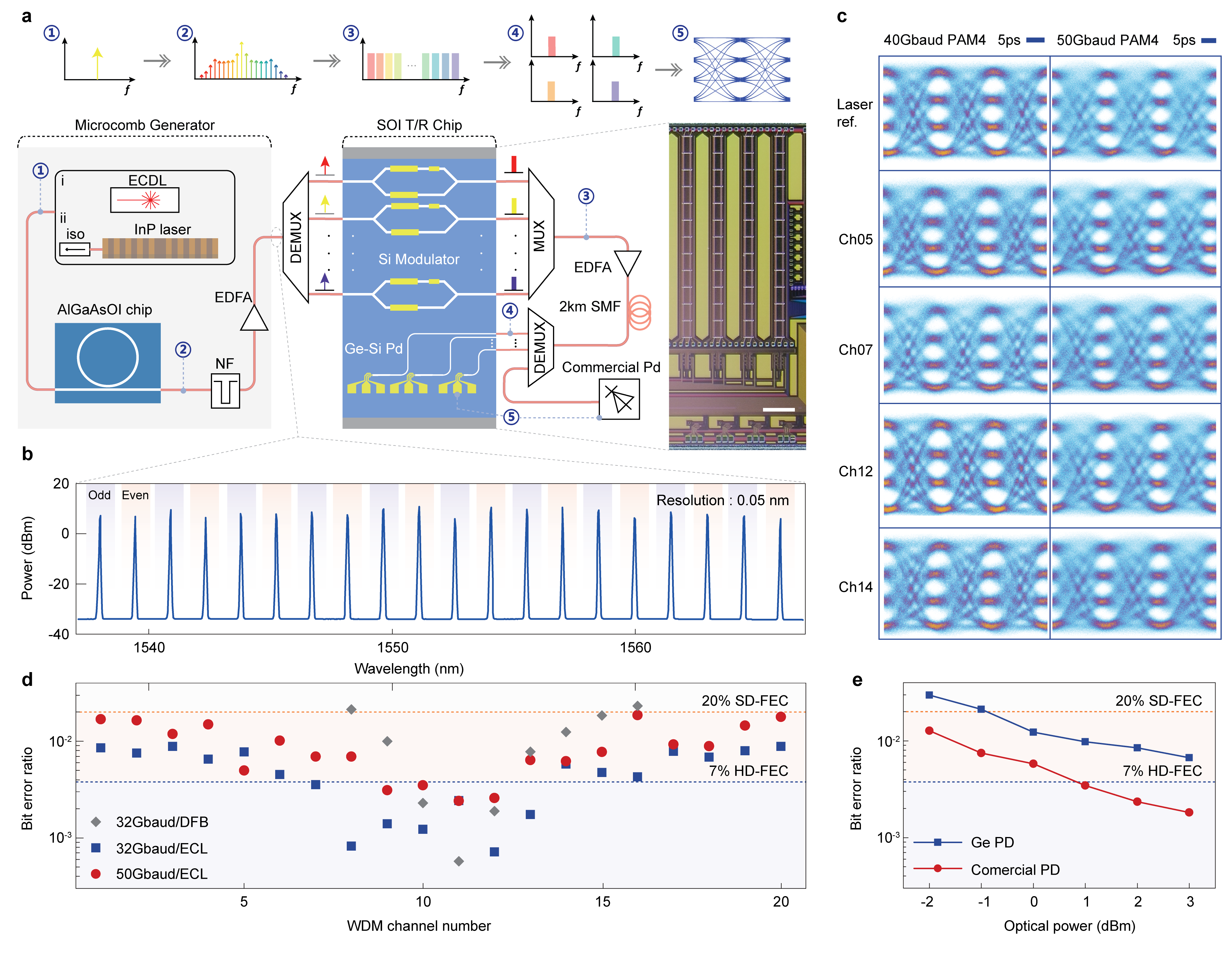}
\caption{\textbf{Transmission results. a}, Schematic of the microcomb-based data transmission set-up. The dark-pulse Kerr comb source is pumped by a c.w. laser, which can be generated by (i) a commercial external cavity diode laser or (ii) a distributed feedback laser chip. The generated comb is then sent into a silicon photonics T/R chip. Labels: ECL, external cavity laser; WSS, wavelength selective switch; EDFA, erbium-doped fibre amplifier; SMF, single mode fibre; PD, photodiode. Scale bar: 500 \textmu m
{\bf b}, 20-line comb spectrum at C band as the multiwavelength source before injecting into the silicon photonics T/R chip. The central three comb lines are attenuated by a notch filter in order to flatten the spectrum.
{\bf c}, Typical eye diagrams of the chosen channel after modulation by silicon photonics modulators at different symbol rates (32, 40, and 50 Gbaud). 
{\bf d}, BER for each comb line. The blue squares and red circles indicate the ECL pumped comb data transmission results at the symbol rate of 32 Gbaud and 50 Gbaud, respectively. All channels are considered within the given HD-FEC (3.8 $\times$ 10$^{-3}$) or SD-FEC (2 $\times$ 10$^{-2}$) threshold (blue and orange dashed line), which corresponds to an error free transmission data rate of 2 Tbit/s. The grey diamond markers show the performance when pumping the AlGaAs microresonator with a DFB chip. 
{\bf e}, BER vs receiving power comparison between an on-chip Ge-Si PD and a commercial PD with the variation of the receiving power. The main limitation of the Ge-Si PD is the nonoptimized frequency response. (Supplementary Note III) 
}
\label{fig3}
\end{figure*}

\vspace{3pt}
\noindent 
\maketitle
\textbf{Silicon photonics engines}\\
\noindent A monolithic silicon photonics circuit is used to process the generated comb lines for diverse optoelectronics systems. Such ‘silicon photonic engines’ provide functionality such as filtering, modulation, multiplexing, time delay and detection on the same chip. The SiPh devices used here are fabricated by a commercial SiPh foundry in a one-to-one 200 mm wafer run with its standard 90 nm lithography SOI process. The waveguide loss in this SiPh platform is approximately 1.2 dB/cm at the C-band. More details can be found in Methods.

Fig. \ref{fig2}e-j show essential photonic building blocks of the optical processing engines and their key performance metrics. For signal encoding, MZI travelling-wave PN depletion modulators with $>$33 GHz EO bandwidth are employed (see Fig. \ref{fig2}e). Heaters are used to match up the modulators with the comb channels by thermal tuning (see Fig. \ref{fig2}f). A representative result for such phase compensation in a modulator at different channel wavelengths is shown in the left panel of Fig. \ref{fig2}e. Meanwhile, a deep trench process is utilized to isolate each micro heater in order to diminish thermal cross talk. To implement on-chip true time delays, spiral waveguides with adiabatic bends are designed, as shown in Fig. \ref{fig2}g. The deviation of a 60 ps silicon delay line used for the microcomb based system in this paper is within 3 ps. Fig. \ref{fig2}h shows the vertical epitaxial Ge photodetector (PD) with 0.5$\sim$0.8 A/W at different on-chip power levels, and with a saturation power of approximately 20 mW. To control the comb lines individually, the wavelength selective switch used here is a microring filter array, as shown in Fig. \ref{fig2}h. A 180 GHz-wide (2 FSR) channel selecting range can be obtained with 20 mW heater power. More details of the coupling and monitoring devices can be found in Supplementary Note III. In addition, the SiPh devices support system-level assembly with electronic integrated chips (EICs) for signal amplification (See Fig. \ref{fig2}j), allowing future seamless integration of low noise trans-impedance amplifiers and high-speed drivers.

\vspace{6pt}
\noindent
\large\textbf{System demonstrations}\\
\normalsize
\noindent Here, we illustrate the potential of bridging these two fields of integrated photonic technology by presenting two pivotal system-level demonstrations: (1) the first microcomb-based integrated photonic data link for optical communications with greatly increased data rate compared to traditional Si-based transceivers and (2) a rapidly reconfigurable microcomb-based microwave photonics filter with unprecedented integration level, where both the source and the processing elements are realized via integrated photonic technologies.

\vspace{3pt}
\noindent
\textbf{Parallel optical data link}
\noindent\\
A PAM-4 WDM transmission experiment is performed here with a combination of AlGaAsOI dark-pulse Kerr comb source and silicon photonics T/R chips. The schematic diagram of the transmission system is shown in Fig. \ref{fig3}a. Considering both OSNR and bandwidth requirement for advanced modulation formats (i.e. 50 Gbaud PAM-4), we select a dark-pulse comb with 2-FSR spacing ($\sim$180 GHz) as the WDM source. For pumps, a DFB laser chip and a commercially available external cavity laser pumped source are used respectively, with equal on-chip pump power of $\sim$10 mW. In previous microcomb-based communication systems \cite{OC1,OC4}, an extra line-by-line shaping process has always been necessary to flatten the comb power within the operation band, which introduces extra system complexity and power consumption on the transmitting side. In this work, thanks to the strong thermal effect, the AMX strength of the AlGaAs microresonator can be thermally pre-set in order to obtain a coherent microcomb with less disparate power distribution across the operation band. Thus, only a notch filter is required to attenuate the central three comb lines for the subsequent equalized comb amplification. The spectrum, after amplification, is shown in Fig. \ref{fig3}b, in which 20 consecutive comb modes (from 1537 to 1567 nm, $\sim$3.75 THz wide) are displayed with $<$ 5dB power difference. A simplified scheme is used to verify the chip-scale data transmission capability for carrying multi-Tbit/s. The comb lines are filtered out and split into odd/even test bands by a wavelength selective switch (WSS) and afterwards launched into the silicon photonics T/R chip, which includes the Si modulators and Ge photodiodes. On each WDM channel the SiPh modulators encode the carrier into PAM-4 signal format at symbol rates from 32 to 50 Gbaud. Fig. \ref{fig3}c shows representative examples of eye diagrams after traversing 2 km fibre links. At the receiving side, the signal is partly coupled to an on-chip Ge photodiode, while the remaining part is sent into a commercial PD for performance comparison. The optimized receiving power for each channel is 2$\sim$3 dBm. The bit-error ratio (BER) of each channel is calculated after direct detection (see Methods). 

Such a DWDM scheme can greatly improve the aggregate bit rate carried by the SiPh data transmitter while maintaining excellent scalability. In our proof-of-concept demonstrations, 20 comb lines in C-band are used as the source.  Fig. \ref{fig3}d shows the BERs result under three scenarios: (i) 32Gbaud and (ii) 50Gbaud PAM4 with ECL pump, (iii) 32Gbaud PAM-4 with DFB pump. Considering the ECL pumped microcomb, 7 (4) channels are below the 7\% hard-decision forward error correction (HD-FEC) threshold at the symbol rate of 32 (50) Gbaud, with the remaining channels below the 20\% soft-decision forward error correction (SD-FEC) threshold. In this case, the microcomb-based SiPh transmitter enables a baud rate of 50 Gbaud per single lane, corresponding to an aggregate bit rate of 2 Tbit/s (1.65 Tbit/s for net rate after FEC overhead subtraction). The wavelength dependent BERs mainly result from the increased noise of the pre-amplifier at the edge of its operation band. For a higher-level integrated system, the commercial ECL pump is replaced by a DFB laser chip. An optical isolator is deployed between the DFB laser and AlGaAsOI microresonator to eliminate the reflection. With the integrated pump source, the transmitter achieves a total data transmission rate of 448 Gbit/s, with 7 channels under the given FEC threshold. Another advantage of SiPh is the possibility of integrating the transmitter and receiver. BER results after O/E conversion by both commercial III-V photodiodes and on-chip Ge photodiodes are shown in Fig. \ref{fig3}e. At the 20\% SD-FEC threshold, the penalty between two devices is approximately 2.3 dB at 32 Gbaud, which may result from the bandwidth limitation of the non-optimized structure in the Ge absorption region. 

\begin{figure*}[ht]
\centering
\captionsetup{singlelinecheck=no, justification = RaggedRight}
\includegraphics[width=18cm]{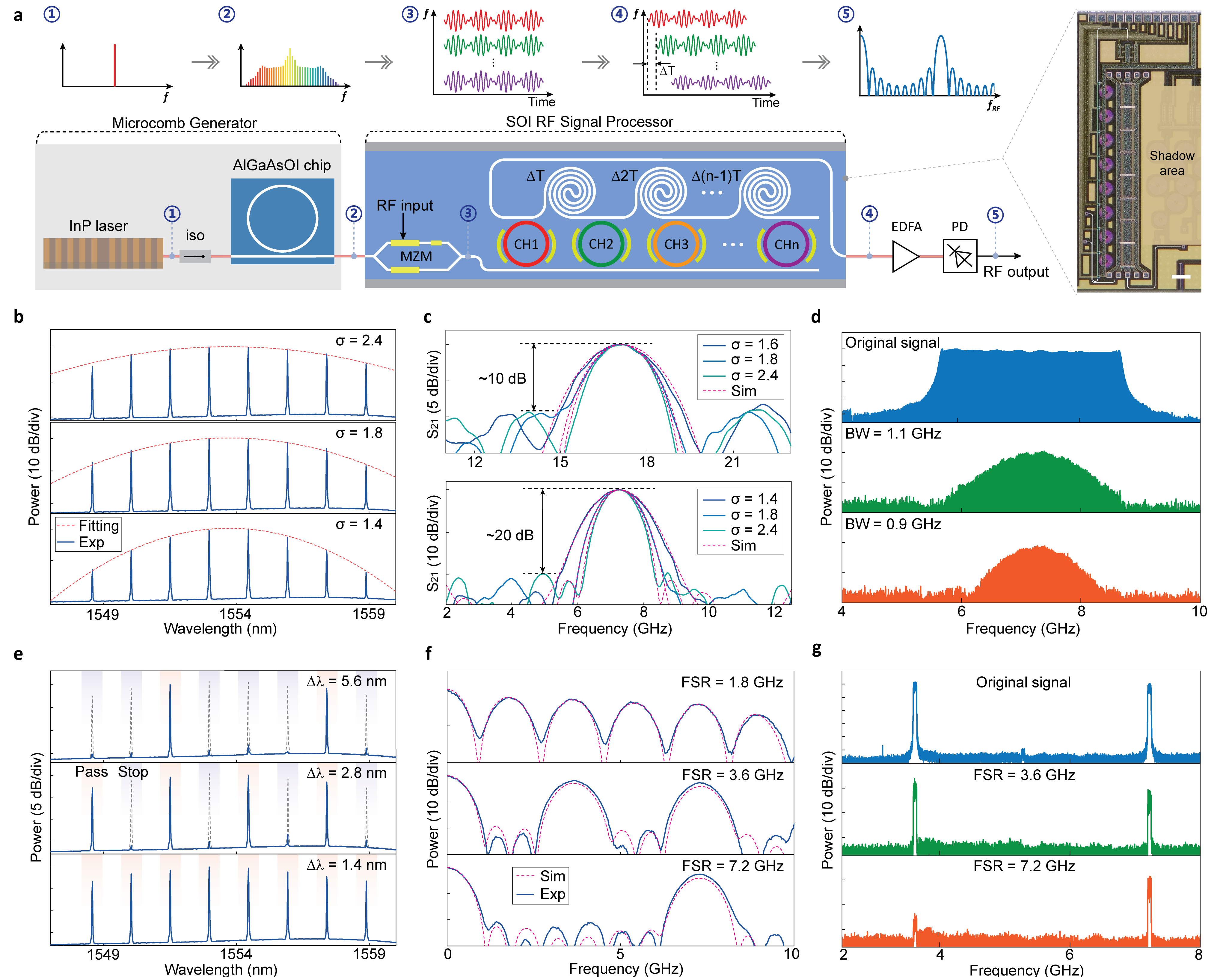}
\caption{\textbf{Reconfigurable MPF results. a}, Schematic diagram of the setup to perform microcomb-based reconfigurable MPF. The setup mainly consists of a microcomb generator based on a DFB-pumped AlGaAs microresonator, a chip-scale silicon photonic RF signal processor for RF signal broadcasting, comb lines shaping \& delay, and an off-chip fast PD. Scale bar: 200 \textmu m
{\bf b}, Optical spectra of Gaussian-apodization comb lines for bandwidth programming ($\sigma$: Gaussian factor; blue: experiment; red: Gaussian fitting).
{\bf c}, RF filtering responses of the MPF with various passband bandwidth, based on the setup $\#$1 (upper panel) and setup $\#$2 (lower panel). Setup $\#$1: The delays between comb lines are produced by on-chip spiral delay paths, as with the configuration in (a). Setup $\#$2: The delays are produced by dispersive propagation from a spool of SMF, see Supplementary Note V. 
{\bf d}, Proof-of-concept demonstration of RF filtering of a wideband RF signal. From top to bottom: RF spectra of original signal, signal after 1.1 GHz BW filter, and signal after 0.9 GHz BW filter.  
{\bf e,f}, Optical spectra (e) and corresponding RF responses (f) of the MPF with various FSRs, produced by modifying the comb line spacing. $\Delta \lambda$ : wavelength distance between adjacent comb lines.
{\bf g}, Proof-of-concept demonstration on RF filtering of a complex dual-channel RF signal. From top to bottom: RF spectra of original signal, signal after 3.6 GHz FSR filter, and signal after 7.2 GHz FSR filter.
}
\label{fig4}
\end{figure*}

\vspace{3pt}
\noindent
\textbf{Reconfigurable microwave photonic filter} \\
\noindent Based on the combination of AlGaAs dark-pulse microcombs and a silicon photonic signal processor, a reconfigurable microwave photonic filter (MPF) is constructed using a tapped delay line (TDL) scheme \cite{RFPart1}. It is worthwhile to mention that TDL-based MPFs can follow two approaches depending on whether the tap delays are produced by non-dispersive (true-time) delay lines \cite{RFPart2,RFPart3} or dispersive delay lines \cite{RFPart4,RFPart5,RFPart6,RFPart7,RFPart8,RFPart9}. In this work, both approaches are implemented. The schematic diagram for true-time delay setup is shown in Fig. \ref{fig4}a. As in the data link DFB test, a chip-based DFB laser is used to pump the dark-pulse microcomb. Combs with 2-FSR spacing (180 GHz) are generated and served as taps for the MPF. The comb lines are then manipulated by a silicon photonic signal processor containing a high-speed MZM, an 8-channel add-drop microring array (MRA) and spiral delay lines. The input RF signal is broadcast onto comb lines by the MZM. The MRA here acts as an on-chip optical spectral shaper (OSS) for the comb lines, performing spectrum slicing, line-by-line pulse shaping (weighting on taps) and spectrum recombination in sequence. A cluster of spiral waveguides are inserted to obtain a fixed time delay ($\Delta$T) between adjacent taps. Finally, the processed comb lines are beaten in an off-chip fast PD to synthesize the RF filtering profiles. More details about the experimental setup are described in Methods. Because the nonuniformity of delays due to the inevitable fabrication errors will degrade the filtering performance, the second TDL-MPF approach is also implemented to further determine the optimal filtering performance: a spool of SMF is used instead of the on-chip spiral delay lines to produce dispersive delay (see in Supplementary Note V). 

%Fig4

The system shows flexible reconstruction features in terms of passband bandwidth (BW) and RF FSR. Fig. \ref{fig4}b depicts the optical spectra using Gaussian apodization on comb lines for passband BW reconfigurability \cite{RFPart4}. The corresponding RF filtering responses are given in the two panels in Fig. \ref{fig4}c; they are measured in the non-dispersive delay (upper panel) and dispersive delay (lower panel) configuration, respectively. The 3-dB bandwidth of the MPF in non-dispersive delay way can be continuously adjusted within a range of 1.97-2.42 GHz by tuning the Gaussian parameter $\sigma$ from 2.4 to 1.6. The main-to-sidelobe suppression ratio (MSSR) is about 10 dB. Better performance is achieved using the dispersive delay scheme, for which all filtering profiles have a MSSR higher than 20 dB, and the 3-dB bandwidth is tunable at the sub-gigahertz level, from 0.95 to 1.35 GHz. The results in Fig. \ref{fig4}e-f, measured under the dispersive delay scheme, show the reconfigurability of RF FSR by modifying the comb line spacing: Comb line spacings of 5.6 nm, 2.8 nm and 1.4 nm result in RF filtering response FSRs of 1.8 GHz, 3.6 GHz and 7.2 GHz, respectively. The similar FSR multiplication of MPF has been reported previously and explained by temporal Talbot effects \cite{RFPart5}. However, in ref. \cite{RFPart5}, a more complex Talbot processor is based on discrete devices, which approach increases the power dissipation and makes the system less stable. In contrast with other state-of-the-art microcomb-based MPFs using either bulk OSS \cite{RFPart6,RFPart7} or changing soliton states \cite{RFPart8}, this work significantly advances the degree of integration and the reconfiguration speed ($\sim$53 \textmu s, see Supplementary Note VI), which are crucial for modern wireless communications and avionic applications.

As a paradigm demonstration toward real-world applications, RF filtering on a practical microwave signal is illustrated in Fig. \ref{fig4}d and Fig. \ref{fig4}g. First, an ultra-wideband RF signal covering from 5.5 GHz to 9 GHz is generated as input for the MPF, as shown in Fig. \ref{fig4}d. By adjusting the Gaussian apodization factor, the BW of this MPF can be reconfigured from 0.9 GHz to 1.1 GHz to allow various passband widths. Moreover, to validate the FSR reconfigurability, a complex RF test signal is prepared with a 50 Mb/s QPSK modulation at 3.6 GHz and a 50 Mb/s QPSK modulation at 7.2 GHz (Fig. \ref{fig4}g). It can be observed that the QPSK modulation spectra at 3.6 GHz and 7.2 GHz will both pass through when the FSR is set to 3.6 GHz. On the other hand, if the FSR is reconfigured to 7.2 GHz, only the QPSK spectra at 7.2 GHz will be selected and the QPSK spectra at 3.6 GHz is almost completely rejected.

\vspace{6pt}
\noindent
\large\textbf{Discussion} \\
\normalsize
\noindent The performance of these systems can be further improved by optimizing the integrated devices or employing superior signal processing techniques. In our communication experiment, the total data rate is lower than that in previous reports for coherent long-haul transmission \cite{OC1,OC2,OC4,OC5}, where the Kerr microresonator is the only integrated device in the whole set-up. Additional multiplexing techniques (i.e. SDM, PDM) and higher modulation formats (i.e. PAM6, PAM8) could be used to boost the transmission capacity. Meanwhile, only 20 comb lines in C band are used due to the bandwidth limitation of the EDFA. We infer that the data rate can be further scaled up to $>$10 Tbps by broadening the operation wavelength to L and S band or introducing more complex multiplexing methods. The performance of the DFB pumped integrated comb source is mainly limited by the relatively high noise floor of the free running DFB laser (see Supplementary Note III), which lowers the OSNR; performance could be improved by employing a narrow bandwidth filter after the pump or an on-chip optical filter for comb distillation \cite{Dtl1}. Moreover, by employing a power-optimized scheme for comb generation and amplification, up to one order of magnitude power consumption reduction can be obtained on the source side compared with the state-of-art integrated tunable laser assemblies \cite{OC2}. It is also worth noting that the 180 GHz comb frequency spacing in our work supports the state-of-art single lane 200G silicon photonics data transmission \cite{200GSiPh}, which is a promising solution for future data-center interconnects \cite{180GPAM4}. For the RF filter, narrower filtering bandwidth (down to sub Gigahertz) and higher tuning resolution can be obtained by increasing the number of tap channels used in the FIR configurations \cite{RFPart1}, i.e. expanding of the microring array. Moreover, the off-chip fiber based dispersive delay may be replaced by a novel design, such as a photonic crystal waveguide \cite{RFPart9}, for higher level integration.

We expect that more integrated functionality will be incorporated in the future, culminating in fully integrated microcomb-based optoelectronic systems. For instance, self-injection locked dark-pulse microcomb sources \cite{SIL1,DP_SIL1} could be monolithically realized by using heterogeneously integrated III-V lasers and microresonators \cite{SIL3}. The integration of AlGaAsOI microcombs and semiconductor optical amplifiers (SOAs) could supply sufficient power to replace the discrete EDFAs, thereby enabling direct coupling between III-V source chips and silicon chips by butt coupling or photonic wire bonding \cite{PWB1,PWB2}. Furthermore, photonic SOI elements can be combined with application-specific electronic circuits, which will further improve the compactness and power-efficiency. Considering the versatility offered by the technologies, the marriage between microcomb and foundry-based silicon photonics demonstrated here provide a mass-produced and low-cost solution to a broad range of optoelectronics applications, therefore facilitating the next generation of integrated photonic.

\vspace{6pt}
\noindent \textbf{Methods}\\
\begin{footnotesize}
\noindent \textbf{Design and fabrication of the devices. } 
The ring waveguides of the AlGaAsOI resonators were designed to work within the normal dispersion regime at C band, with dimensions of 400 nm $\times$ 1000 nm. The width of the bus waveguide at the facet was designed to be 200 nm for efficient chip-to-fiber coupling. The fabrication of AlGaAs microresonators was based on heterogeneous wafer bonding technology. The epitaxial wafer growth was accomplished using molecular-beam epitaxy (MBE). A 248 nm deep-ultraviolet (DUV) stepper was used for the lithography. A photoresist reflow process and an optimized dry etch process were applied in waveguide patterning to minimize waveguide scattering loss. More fabrication details can be found in ref. \cite{AlGaAs1} and \cite{AlGaAs2}. The silicon photonics PIC, including its silicon modulators and Si-Ge PDs, was fabricated on a 200 mm SOI wafer with a silicon-layer thickness of 220 nm and a BOX layer thickness of 2 \textmu m using CMOS-compatible processes at CompoundTek Pte. Both the modulator and PD exhibit 3-dB bandwidths in excess of 30 GHz (see Supplementary Note III). In our experiment, lensed fibers with different mode field diameter (MFD) were selected for the AlGaAsOI and SOI chips; the coupling loss is 3-5 dB per facet for AlGaAsOI waveguides and 2-3 dB per facet for Si waveguides.

\vspace{3pt}
\noindent\textbf{Data transmission experiments.}
In our experiment, the microcomb is first pumped by a commercial tunable laser (Toptica CTL 1550), then by a DFB laser chip for a higher degree of integration. When tuning the pumping wavelength from the blue side to a certain detuned value at around 1552.5 nm, both configurations generate dark-pulses with 2-FSR comb spacing. The detailed experimental setup for data transmission is shown in Supplementary Fig. 5Sb. The comb is amplified by an erbium-doped fibre amplifier (EDFA) and then split into odd/even test bands by a wavelength-selective switch (Finisar Waveshaper 4000s). A silicon modulator and a lithium niobate (LN) modulator (EOspace, 35 GHz bandwidth) are deployed at the odd and even bands, respectively. 10 comb lines in each test band are simultaneously modulated. The modulators are driven at a 32/50 Gbaud symbol rate. The differential PAM-4 signal is generated by a commercial pulse pattern generator (Anritsu PAM4 PPG MU196020A). The insertion loss of the SiPh (LN) modulator is 13 (8) dB. The SiPh modulator undergoes a relatively high loss (including the edge coupling loss of $\sim$2 dB/facet), which results in a power difference between the two test bands. The modulated test bands are then combined by a 50:50 power coupler and launched into another WSS for comb power equalization. At the receiving side each WDM channel encoded by the silicon modulator is sequentially filtered out and measured. Eye diagrams are produced by a sampling oscilloscope (Anritsu MP 2110A) with a 13-tap TDECQ equalizer (see Supplementary Note IV). The BERs are measured on-line by an error detector (Anritsu PAM4 ED MU196040B) with 1dB low frequency equalization and a decision-feedback equalization. 

It is worth noting that the performance is underestimated. In our proof-of-concept test configuration, 10 channels in each test band are modulated at the same time. Considering two photon absorption (TPA) in silicon waveguides, the maximum input power for the silicon modulator is $\sim$13 dBm, which results in only 3 dBm optical power per single lane. Moreover, taking into account the extra penalty introduced by the WSS for power equalization, unnecessary in real-word transmission scenarios, the OSNR for each channel can be at least 10 dB higher. Thus, a better transmission result is attainable.

\vspace{3pt}
\noindent 
\textbf{RF filter experiments.} The DFB-driven dark-pulse Kerr comb exhibits 2-FSR (180 GHz) comb spacing. The initial comb source is amplified by an EDFA, and 8 comb lines in 1547-1560 nm are selected using an optical bandpass filter before injecting into a silicon photonics signal processor chip. The input and output coupling are achieved via grating couplers of $\sim$40\% coupling efficiency. Frequency-swept RF signals with 9 dBm power from a VNA (Keysight N5247A) are applied to the silicon MZM in double-sideband (DSB) format. The tap weighting coefficients are set by adjusting the relative detuning among the comb lines and their corresponding resonance wavelengths in the silicon MRA with TiN microheaters placed on the waveguides. The output light of the silicon chip is split by a 10:90 optical power coupler: 10\% of the light is sent into an optical spectrum analyzer (Yokogawa AQ6370C) for spectral monitoring, while the other 90\% of the light propagates through the follow-up optical link. In the dispersive delay scheme, a spool of 5 km SMF is used to acquire the dispersive delay between adjacent comb lines (taps). Finally, the processed comb lines are beat in a 50 GHz PD (Finisar 2150R) to convert the optical signal into electrical domain. A low-noise EDFA is placed before the PD to compensate for the link insertion loss and coupling loss. 

For the practical demonstrations of RF signal filtering, a 50 GSa/s arbitrary waveform generator (AWG, Tektronix AWG70001) is employed to produce the desired RF input signals. To validate the BW reconfigurability of this filter, an ultra-wideband RF signal is generated, spanning from 5.5 GHz to 9 GHz. To validate the FSR reconfigurability of this filter, a complex RF signal is produced which contains a 50 Mb/s QPSK spectrum modulated at 3.6 GHz and a 50 Mb/s QPSK spectrum modulated at 7.2 GHz. The RF outputs from the AWG are amplified by a linear electrical driver (SHF 807C) before routing to the silicon MZM. The filtered RF signals are detected by a signal analyzer (Keysight N9010B) for spectrum measurement.

\vspace{6pt}
\noindent\textbf{Data availability}\\
The data that supports the plots within this paper and other findings of this study are available from the corresponding authors upon reasonable request. 

\vspace{6pt}
\noindent\textbf{Code availability}\\
The codes that support the findings of this study are available from the corresponding authors upon reasonable request.

\end{footnotesize}
\vspace{20pt}

%_______REFERENCE____________%
\bibliography{Ref.bib}
% \bibliographystyle{naturemag}
% \bibliographystyle{naturesaa}

%%_______REFERENCE_______%%

%--------------------------------------------------------------------------
\vspace{12pt}
\begin{footnotesize}

\vspace{6pt}
\noindent \textbf{Acknowledgment}

\noindent 
The authors thank Shenzhen PhotonX Technology Co., Ltd., for laser packaging support and Theodore J Morin for helpful commentary on the manuscript. The UCSB nano-fabrication facility was used.

\vspace{6pt}
\noindent \textbf{Author contributions}

\noindent 
 The experiments were conceived by H.S., L.C., Y.T. and B.S. The devices were designed by H.S., L.C., and Y.T. The microcomb simulation and modelling is conducted by B.S. The system-level experiments are performed by H.S., Y.T., with the assistance from L.C., B.S., M.J., Z.T., X.Z., Q.J., R.C. and B.B. The AlGaAsOI microresonators are fabricated by W.X. and L.C. The results are analyzed by H.S., Y.T., B.S. and A.N. All authors participated in writing the manuscript. The project was coordinated by H.S. and L.C. under the supervision of S.Y., X.W. and J.E.B. 

\vspace{6pt}
\noindent
\textbf{Additional information} 

\noindent Supplementary information is available in the online version of the paper. Reprints and permissions information is available online. Correspondence and requests for materials should be addressed to X.W. and J.E.B.

\vspace{6pt}
\noindent \textbf{Competing financial interests} 

\noindent The authors declare no competing financial interests.
\end{footnotesize}
%--------------------------------------------------------------------------

\end{document}